# Nouvelles modalités d'électrophysiologie corticale, perspectives en recherche médicale et en physiologie humaine

**Pierre BOURDILLON**[1,2,3], **Linnea EVANSON**[3,4]

*Neurochirurgien titulaire à l'Hôpital Fondation Adolphe de Rothschild, Doctorante à l'Ecole Nomale Supérieure*

## Résumé

Les progrès récents dans la conception de matériaux et en micro et nano-électronique ont modifié en profondeur la conception des électrodes d'électrophysiologie intracrânienne. Il devient possible de produire des électrodes enregistrant l'activité du cortex à une échelle qui n'était jusqu'à présent pas disponible. Cette échelle correspond à l'enregistrement des structures fonctionnelles de base, supposées être les plus élémentaires, utilisée par le cerveau pour son fonctionnement et, au sein de ces structures, de l'activité des différents types de neurones les composant. Après une synthèse concernant les différents types de nouvelles électrodes, est exposée ici une de leurs toutes premières applications pour la description de mécanismes conduisant à la génération d'une crise comitiale focale chez l'épileptique. S'en suit une réflexion plus large sur les perspectives de leur utilisation.

## Mots-clés

PEDOT ; laminaire, électrode ; cortex ; épilepsie ; électrocorticographie ; intracrânien ; silicone ; CMOS

# New modalities of cortical electrophysiology, perspectives in medical research and human physiology

## Abstract

Recent advances in material technology and in micro- and nano-electronics have profoundly changed the design of intracranial electrophysiology electrodes. It is now possible to manufacture electrodes that record cortical activity at a spatial resolution that was previously unthinkable. This high spatial resolution enables recording of the functional structures of the brain, and differentiation of the activity of the different types of neurons composing them. In this paper, we present a review of the different types of electrodes now available, and then suggest one of the first applications for such high resolution electrodes, namely a means to better characterise the mechanisms that generate focal seizures in epileptics. Finally, we reflect more broadly on prospects for their future use.

## Keywords

PEDOT ; laminar, electrode ; cortex ; epilepsy ; electrocorticography ; intracranial ; silicone ; CMOS

---

[1] *Integrative Neuroscience & Cognition Center, Université Paris-Cité, Paris France*
[2] *Cortical Physiology Laboratory, Massachusetts General Hospital, Harvard Medical School*
[3] *Département de Neurochirurgie, Hôpital Fondation Adolphe de Rothschild, Paris France*
[4] *Département d'Etudes Cognitives, Ecole Normale Supérieure, Paris France*

# Introduction

La place de l'électricité au sein du vivant remonte à la description même du concept moderne d'électricité, notamment avec les travaux de Giovani Aldini[1,2] s'appuyant sur ceux de Luigi Galvani et d'Alessandro Volta, la décrivant comme distinct de celle de la foudre, de nature statique. Dans les années et même le siècle suivant ces premières observations et interprétations, la recherche dut essentiellement se baser sur la visualisation de la conséquence des stimulations électriques, l'enregistrement des courants électriques étant alors difficile. Une fois rendu possible, les nerfs, voies de propagation des signaux électriques, furent les premiers étudiés et il fallut attendre la fin du XIX[e] siècle et les travaux de Richard Caton pour pouvoir enregistrer, par le biais d'électrodes corticales, l'activité du cerveau animal[3]. Ce fut une avancée technique, permettant l'amplification du signal et donc son recueil à travers le crâne et le scalp, qui permit les premiers enregistrements de l'activité corticale humaine par Hans Berger en 1929[4] lors de la description de l'électroencéphalographie. Une des étapes essentielles dans le progrès de la compréhension de la physiologie du cerveau humain sur les bases de l'analyse de son activité électrique fut le développement d'électrodes d'enregistrement intracérébrales, permettant son enregistrement de manière continue sur plusieurs jours. Ce développement fut induit par la nécessité de disposer d'enregistrements précis des crises d'épilepsie pour en proposer un traitement chirurgical efficace. Ces électrodes de nature métallique, initialement en acier et désormais faites d'un alliage d'iridium et de platine, enregistrent les potentiels de champs locaux à savoir l'activité globale d'une large population de plusieurs milliers de neurones[5]. S'il n'est donc pas possible d'avoir d'informations quant à l'organisation fonctionnelle des neurones les uns avec les autres, il est en revanche possible de comparer la temporalité de l'activation de plusieurs sites d'enregistrement dans le cerveau et d'ainsi décrire l'organisation et la dynamique des réseaux fonctionnels à une échelle macroscopique. Des enregistrements à l'échelle d'un neurone unique, appelés enregistrements unitaires, ont par ailleurs été réalisés par la miniaturisation des électrodes et l'augmentation de la fréquence d'échantillonnage des amplificateurs, mais ne permirent qu'une étude limitée des interactions entre les neurones, bases essentielles du fonctionnement cérébral. Depuis le développement de cette approche au début de la seconde moitié du XX[e] siècle, peu de progrès ont été réalisés et il faut attendre le début XXI[e] siècle et le développement de la micro puis nano-électronique ainsi que des matériaux compliants pour envisager l'enregistrement du cortex cérébral à différentes échelles. L'objet du présent travail est de présenter ces nouvelles approches et d'aborder les perspectives de leur utilisation.

## Les nouvelles modalités d'enregistrement cortical

### Bases technologiques

La conception des électrodes intracrâniennes de surface ou pénétrant le cortex repose actuellement sur une architecture géométrique et des propriétés physiques mécaniques ne permettant pas un échantillonnage à haute résolution spatiale et exposant à des problèmes d'interface entre l'électrode et le tissu nerveux[6-8]. Pour résoudre ces deux limites propres aux propriétés physiques des électrodes, de nouvelles approches ont été utilisées afin d'obtenir des systèmes d'enregistrement à haute densité spatiale et de grande compatibilité biophysique avec le tissu nerveux. Le principal obstacle était la différence d'ordre de grandeur du module de Young, c'est-à-dire la constante qui relie la contrainte de compression et le début de la déformation d'un matériau élastique isotrope, entre le tungstène (250 GPa), l'iridium-platine (200 GPa), le silicone (150 GPa) et le tissu nerveux (1 à 10 kPa). Cela expose en effet à des lésions tissulaires pouvant conduire à des réactions inflammatoires et la production de gliose réactionnelle pouvant causer une mauvaise interface entre le cerveau et l'électrode et ainsi gêner le recueil de l'activité électrique et impacter le fonctionnement normal du tissu nerveux[6,9]. Le second obstacle était que le matériau choisi permette une architecture pouvant produire une résolution spatiale suffisamment fine pour permettre l'enregistrement d'activité au niveau du neurone unique et cela sur de larges populations cellulaires (ce qui est appelé l'activité multi-unitaire) tout en assurant une impédance basse[10]. Les principales avancées résultèrent de la nanostructuration des surfaces d'enregistrement et l'utilisation d'un couplage ionique et électronique pour la conduction du signal. À ce jour les matériaux les plus aboutis sont :

- Composite platine-silicone : 0.5 $\Omega cm^2$ impédance à 1 kHz, 300µm de diamètre d'électrode, 35 nm d'épaisseur[11].
- Hydrogels incorporant du poly(3,4-ethylenedioxythiophene):poly(styrenesulfonate) (PEDOT:PSS) : $2\Omega cm^2$ impédance à 1 kHz, $1mm^2$ de taille d'électrode, 200 nm d'épaisseur. L'électrode utilise des nano-fils de dioxyde de titane recouvert d'or (Au-TiO2) : $0,22\ \Omega cm^2$ impédance à 1 kHz, $50 \times 50\ \mu m^2$ de taille d'électrode, 3µm d'épaisseur[12].
- Films conducteurs de polymère polypyrrole/polycaprolactone-block-polytetrahydrofuran-blockpolycaprolactone (PPy/PCTC) : $66\ \Omega cm^2$ impédance à 1 kHz, $0,5mm^2$ de taille d'électrode, 15µm d'épaisseur[13].

**Les différents types d'électrodes**

Ces matériaux sont ensuite utilisés pour concevoir des matrices d'électrodes à haute densité permettant l'obtention de la résolution spatiale évoquée au début de cet article. Ces électrodes peuvent être des électrodes de surface, souples, ou bien des électrodes pénétrant le cortex et permettant son enregistrement couche par couche (électrode laminaire). Les électrodes plus significatives et ayant fait l'objet d'application in vivo sont (voir **Figure 1**) :

Neural-Matrix : une matrice d'électrodes pouvant dépasser le millier d'électrodes d'enregistrement (1008) avec des contacts d'enregistrement de $195 \times 270\ \mu m^2$ espacés de 250 à 330 µm sur une matrice de 47,5 µm d'épaisseur de polyamide–élastomère. La densité d'électrode est de 1212 par $cm^2$ [14].

Neuro-Grid : S'il ne s'agit pas de la seule électrode utilisant la technologie Pt-Au-PEDOT:PSS, c'est une des premières publiées et la plus largement utilisée[15]. Il s'agit d'une électrode flexible de 4µm d'épaisseur composée de 256 canaux d'enregistrement de 10 µm² espacés de 30 µm et ayant donc une résolution spatiale de 111 000 électrodes par $cm^2$. La répartition optimale des électrodes au sein de la matrice pour les Pt-Au-PEDOT:PSS en terme de densité de taille d'exploration pour obtenir l'information la plus pertinente est un sujet de réflexion actif rendu possible par la facilité de modifier la taille de la matrice et la répartition des électrodes[16].

Neuro-Pixel : contrairement aux deux électrodes précédentes correspondant à des électrodes fines et souples appliquées à la surface du cortex, ce troisième type d'électrode correspond à une électrode laminaire, à savoir une électrode pénétrant le cortex pour l'enregistrer sur son épaisseur. Contrairement aux générations précédentes ne pouvant enregistrer qu'à l'extrémité de parties performantes (comme la Utah Array n'enregistrant que le troisième feuillet cortical), les électrodes d'enregistrement sont réparties sur toute la longueur, à l'image des électrodes laminaires métalliques[17]. Il s'agit d'une sonde en silicone de 1 cm × 70 µm × 20 µm portant des électrodes (de 250 à 1000) en nitrure de titane de 12 µm² et espacées et 15 µm (impédance de $0,2\ \Omega cm^2$ pour une fréquence d'échantillonnage à 30 kHz).

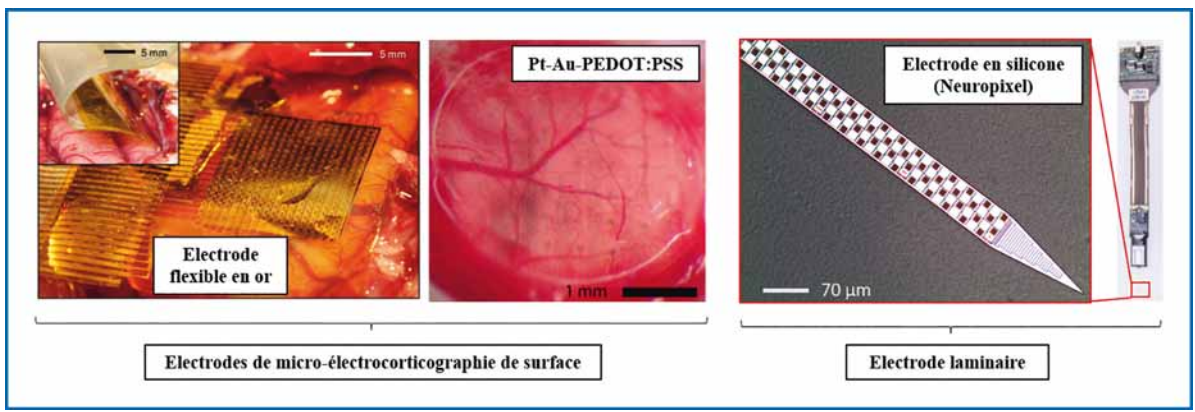

**Figure 1** : exemple d'électrodes de micro-électrocorticographie (en or[14] et Pt-Au-PEDOT:PSS[18]) et laminaire (Neuropixel[19])

Les électrodes transparentes utilisées pour l'optogénétique (principalement basée sur le graphène ou indium tin oxide) et dont la perspective d'application humaine n'est pas immédiate ne sont pas abordées ici.

### Quelles perspectives en recherche médicale, description d'une première approche (travail réalisé dans le cadre du financement post-doctorat Fyssen par le premier auteur)

Les applications de ces nouvelles technologies seront sans nul doute nombreuses dans le domaine de la médecine dans la mesure où des processus pathologiques pourront être observés à une échelle qui jusqu'à présent n'était pas accessible. La première application fut sans surprise une des seules qui était éthiquement acceptable chez le primate humain pour une première approche : lorsqu'il était nécessaire de faire un enregistrement du cortex à l'aide des électrodes cliniques utilisées en routine. Il s'agit de la situation où un patient doit avoir une chirurgie dans le cadre du traitement d'une épilepsie résistante au traitement médicamenteux et où l'on cherche à identifier la zone du cerveau responsable de la génération des crises d'épilepsie. L'utilisation d'une nouvelle électrode en plus de celle utilisée dans le cadre du soin ne rajoute qu'un risque minime pour le patient et a été considérée comme acceptable par le comité de l'éthique en charge de la recherche (Massachusetts General Hospital, Harvard Medical School). C'est ainsi que chez 10 patients, une ou plusieurs électrodes de type laminaire ont été implantées et que 30 crises d'épilepsie ont été enregistrées. Certaines électrodes étaient situées en dehors de la zone de génération des crises d'épilepsie (mais dans des zones pouvant être impliquées dans leur diffusion) et d'autres en leur sein. Il fut donc possible de réaliser une comparaison en fonction du site d'implantation. L'analyse de ces données a permis de mettre en évidence une origine purement infra-granulaire et granulaire (soit les couches les plus profondes du néocortex) comme source de la génération des anomalies à l'origine de la genèse de la crise d'épilepsie (voir **Figure 2**). À l'inverse, dans les zones de propagation des crises, il fut noté que la génération des anomalies épileptiques ne concernait que la couche granulaire, à savoir la couche la plus superficielle du néocortex (*Bourdillon et al.* Article en cours de révision). Ces résultats sont inattendus car l'organisation normale du néocortex suit un schéma vertical très robuste, quelle que soit sa localisation, dans des unités physiologiques appelées colonne corticale. En effet, l'information arrive par la couche granulaire, et est ensuite transmise à la couche infra-granulaire ou un premier traitement de l'information est réalisé avant d'être projeté, via des neurones pyramidaux, vers la couche la plus superficielle du cortex granulaire. Un nouveau traitement de l'information est alors réalisé, notamment comparant l'information reçue avec la prédiction faite localement, et différence entre ces deux éléments est ensuite projetée à distance (vers une autre colonne corticale) par des cellules pyramidales de la couche supra-granulaire (ou parfois plus profonde)[20,21].

Les constatations faites suggèrent donc une déconnexion fonctionnelle de la couche supra-granulaire dans les colonnes corticales au sein du foyer épileptique lors de la survenue d'une crise d'épilepsie. Les électrodes mécaniques utilisées, à travers leurs 24 canaux d'enregistrement ne permettaient cependant pas de faire de l'identification de neurones sur les base de leur activité multi-unitaire. Cependant un nombre plus élevé et une répartition plus dense des électrodes permettrait d'obtenir une telle information (par exemple de suivre l'activité d'interneurones inhibiteurs, de cellules pyramidales, de cellules en miroir de la couche 6...) et d'identifier quels sont les mécanismes responsables de ce puissant découplage vertical au sein d'une colonne. L'électrode Neuropixel possède cette capacité et il a été réalisé au sein du laboratoire les premiers enregistrements chez l'humain. Cependant, seul du signal inter-critique est pour le moment disponible et des obstacles de traitement de signal restent à franchir (en particulier la correction des artefacts de mouvement)[22].

Par ailleurs, l'utilisation d'électrodes laminaires permet d'avoir une vision « verticale » au sein d'une colonne corticale unique mais n'offre pas de possibilité d'analyser les interactions entre les colonnes corticales dans une même unité fonctionnelle ou au sein de la zone épileptogène (voir **figure 3**). Cela est en revanche possible en utilisant des électrode de surface à haute résolution de type Pt-Au-PEDOT :PSS, la résolution spatiale, de l'ordre de 30 à 50 μm selon les configurations, étant inférieur à la taille d'une

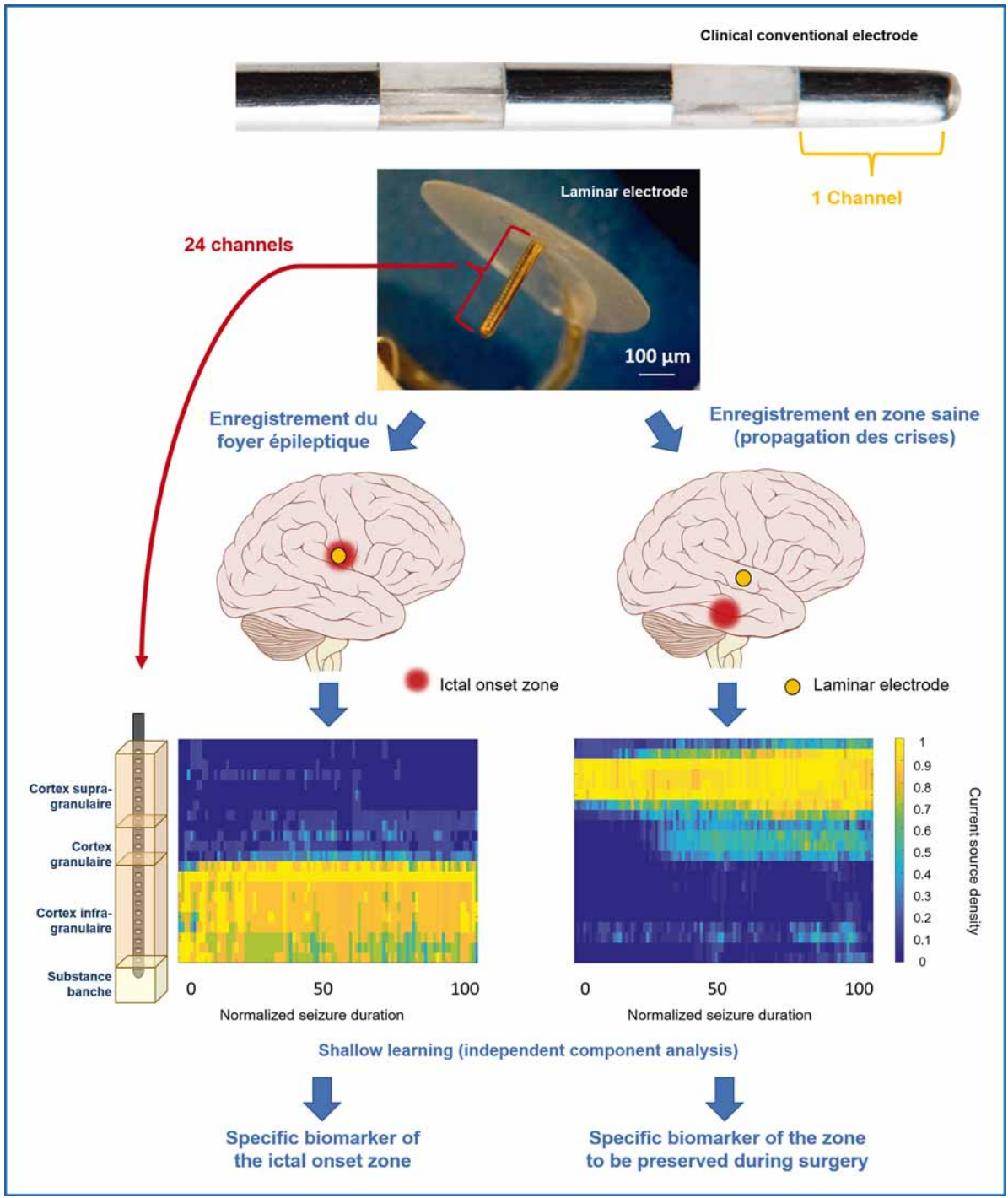

**Figure 2** : Résumé graphique des premiers résultats d'enregistrements laminaires réalisés au sein du foyer de génération des crises ainsi que sur sa voie de propagation. L'analyse du signal repose sur le calcul de la dérivée spatiale secondaire des potentiels de champ permettant l'estimation des flux de courant entre les différentes couches du néocortex. L'utilisation d'une analyse en composante indépendante a permis l'identification d'un biomarqueur spécifique du foyer de génération des crises comitiales.

153

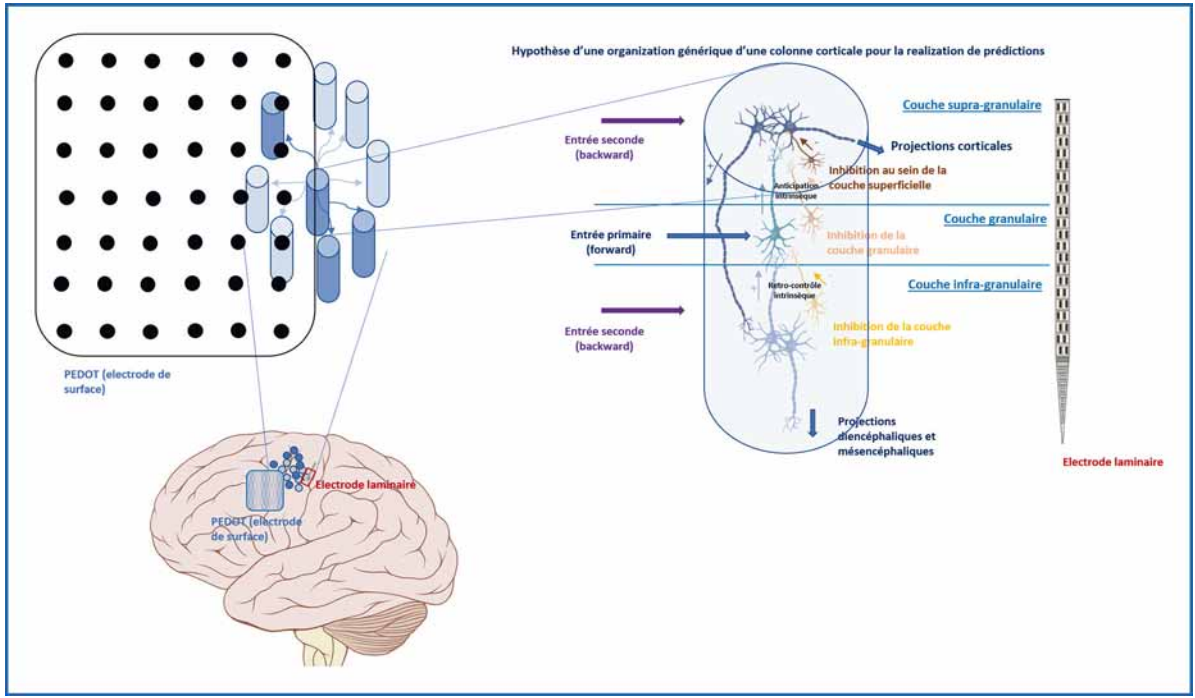

**Figure 3 :** Exemple d'enregistrement d'une part par Pt-Au-PEDOT :PSS permettant une analyse d'une unité fonctionnelle « horizontale » à l'échelle des colonnes corticales, sièges supposés des opérations de calcul élémentaire du néocortex, et d'autre part par une Neuropixel permettant l'enregistrement « vertical » rendant possible au niveau de l'échelle du neurone d'avoir accès aux mécanismes cellulaires impliqués dans la réalisation d'un calcul. L'exemple pris ici testerait le modèle dit du « predictive-coding » proposé par Karl Frinston[20] et prédisant qu'une colonne corticale est responsable de comparer l'information reçue avec une prédiction faite et de transmettre ensuite une quantification de l'erreur constatée. Il s'agirait d'un mécanisme générique de traitement de l'information par le néocortex humain. Cette hypothèse n'a pas encore, à ce jour, pu être testée.

colonne corticale. Il a été réalisé des premiers enregistrements humain à l'aide de ces électrodes au sein du laboratoire chez des patients épileptiques permettant d'obtenir un signal de bonne qualité[16]. Cependant, il n'a pas été à ce jour enregistré de crise d'épilepsie et l'ensemble des données analysées a concerné des enregistrements de l'activité inter-critique, mettant néanmoins en évidence une dynamique pathologique au sein du foyer épileptique[16].

### Perspective pour la compréhension de la physiologie corticale humaine

La possibilité de réaliser des enregistrements corticaux à la fois de surface sur des régions de plusieurs cm$^2$ et du cortex dans sa profondeur avec une résolution inférieure à 50 µm ouvre la voie à la description d'une physiologie corticale encore inconnue chez l'humain. En effet l'utilisation in vivo de ces technologies chez l'animal (principalement, le rongeur, le carnivore et le primate non-humain) ont conduit à des avancées importantes sur la compréhension des mécanismes élémentaires de fonctionnement du cerveau. En effet, jusqu'à présent, les enregistrements reposaient essentiellement sur un enregistrement macroscopique du cerveau et sur la localisation et la temporalité des événements au sein de grands réseaux fonctionnels.

La possibilité d'observation à une échelle plus élémentaire (les neurones eux même au sein de la plus petite unité fonctionnelle supposée au sein de notre cerveau) permet d'envisager la description d'une physiologie corticale encore inconnue chez le primate humain. En effet, il existe pour ces processus élémentaires du fonctionnement cortical une forte variabilité inter-espèce et l'architecture du néocortex humain diffère en de nombreux points de celle des autres espèces et y compris du primate non-humain, en particulier dans l'organisation de la couche 1 et de la couche 6, cette dernière étant jusqu'à ce jour inaccessible aux explorations électrophysiologique. Par ailleurs, le primate humain se caractérise sur le plan cognitif par la possibilité de construire des raisonnements récursifs, ce qui semble indispensable à certaines capacités spécifiques à l'espèce tel la construction du langage, de structures mathématiques ou la métacognition[23-25]. Outre la compréhension et l'étude des hypothèses émises concernant le fonctionnement élémentaire du cerveau comme un calculateur faisant des inférences et quantifiant les erreurs de prédiction[20,23,26] (voir **Figure 3**), il devient possible de réaliser des comparaisons inter-espèces et d'envisager d'isoler le ou l'ensemble de processus permettant à l'espèce humaine de produire des raisonnement récursifs.

> "La possibilité de réaliser des enregistrements corticaux à la fois de surface sur des régions de plusieurs cm² et du cortex dans sa profondeur avec une résolution inférieure à 50 µm ouvre la voie à la description d'une physiologie corticale encore inconnue chez l'humain."

## Conclusion

Le développement de nouveaux matériaux pour la conception d'électrodes permettant l'enregistrement du cerveau à l'échelle des unités fonctionnelles élémentaires que sont les colonnes corticales et les neurones les composant permet d'aborder des questions aux réponses jusqu'ici inaccessibles en terme de physiologie corticale et de physiopathologie. Dans un premier travail nous avons prouvé cela à l'échelle de la question très précise du mécanisme de génération des crises d'épilepsie et avons réussi à isoler un biomarqueur spécifique ouvrant la voie à une application concrète neurochirurgicale à très court terme (guidage électrophysiologique de la résection du foyer épileptique). Par ailleurs il a été possible de montrer la faisabilité des enregistrements avec l'ensemble des techniques proposées. Outre les progrès possibles dans les champs décrits précédemment, il est également possible que ces électrodes permettent un décodage de l'activité neurale avec une grande qualité, et l'utilisation d'électrodes implantées de manière chronique pourraient être utilisées pour de l'interface cerveau machine, notamment dans l'optique de compensation de fonctions altérées dans les suites de lésions du système nerveux central.

## Introduction

The study of electricity in living organisms dates back to the description of the modern concept of electricity, by Giovani Aldini[1,2] whose work was based on those of Luigi Galvani and Alessandro Volta. Aldini described electricity as distinct from lightning and of a static nature. In the years and even the century that followed these first observations and interpretations, research had to be based on the visualization of the consequences of electrical stimulation, as recording electrical currents themselves was infeasible. Once it became possible to record electrical currents, nerves, which can be thought of as voices for the propagation of electrical signals, were the first thing to be studied; however, it was not until the end of the 19th century when the work of Richard Caton made it possible to record the activity of the animal brain[3] for the first time, by means of cortical electrodes. This was a technological advancement, allowing the amplification of neural signal and therefore enabling its detection through the skull and scalp, a technology which in turn allowed the first recordings of human cortical activity by Hans Berger in 1929[4] following Berger's invention of electroencephalography. One of the essential steps in understanding the physiology of the human brain on the basis of the analysis of its electrical activity was the development of intracerebral recording electrodes, allowing recordings to be taken continuously over several days. This development was induced by the need to have precise recordings of epileptic seizures in order to propose effective surgical treatments. These metallic electrodes, initially made of steel and now made of an alloy of iridium and platinum, record the local field potentials, i.e. the overall activity of a large population of several thousand neurons[5]. Thus, these electrodes provide no information about the functional organization of neurons, however, by comparing the temporality of the activation of several recording sites in the brain it is possible to describe the organization and dynamics of functional networks on a macroscopic scale. Recordings at the scale of a single neuron, called unitary recordings, have also been made by miniaturizing the electrodes and increasing the sampling frequency of the amplifiers, but this only allowed a limited study of the interactions between neurons, gleaning information about only the essential bases of cerebral function. After the development of this approach in the second half of the 20th century, little progress was made for over 50 years. Finally, the recent development of micro-then nano-electronics, as well as compliant materials, in the beginning of the 21st century has led to recording of the cerebral cortex at different scales. The purpose of this work is to present these new approaches and to discuss the prospects for their use.

## New modalities of cortical recording

### Technological basics

The design of surface intracranial electrodes or electrodes penetrating the cortex is currently based on a geometric architecture and mechanical physical properties that do not allow high spatial resolution sampling and lead to interface problems between the electrode and the nervous tissue[6-8]. To overcome these two limitations that are specific to the physical properties of the electrodes, new approaches have been used to design recording systems with high spatial density and high biophysical compatibility with nervous tissue. The main obstacle was the difference in the order of magnitude of the Young's modulus, which is the constant that connects the compressive stress and the onset of the deformation of an isotropic elastic material, between tungsten (250 GPa), iridium-platinum (200 GPa), silicone (150 Gpa) and nerve tissue (1 to 10 kPa). This exposes nerve tissue to stress that can lead to inflammatory reactions and the production of reactive gliosis that can cause a poor interface between the brain and the electrode and thus interfere with the collection of electrical activity and impact the normal functioning of the tissue[6-9]. Another requirement was that the chosen material allows an architecture that can produce a sufficiently fine spatial resolution to allow the recording of activity at the level of a single neuron and on large cell populations (what is called multi-unit activity) while ensuring low impedance[10]. The main advances so far have resulted from the nanostructuring of recording surfaces and the use of ionic and electronic coupling for signal conduction. To date, the most successful materials are:

- Platinum-silicone composite: 0.5 $\Omega cm^2$ impedance at 1 kHz, 300 μm electrode diameter, 35 nm thickness[11].
- Hydrogels incorporating poly(3,4-ethylenedioxythiophene):poly(styrenesulfonate) (PEDOT:PSS): 2 $\Omega cm^2$ impedance at 1 kHz, 1mm² electrode size, 200 nm thickness. The electrode uses gold-coated titanium dioxide (Au-TiO2) nanowires: 0.22 $\Omega cm^2$ impedance at 1 kHz, $50 \times 50$ μm² electrode size, 3 μm thickness[12].
- Polypyrrole/polycaprolactone-block-polytetrahydrofuran-blockpolycaprolactone (PPy/PCTC) polymer conductive films: 66 $\Omega cm^2$ impedance at 1 kHz, 0.5mm² electrode size, 15μm thickness[13].

### The different types of electrodes

These materials are used to design high-density electrode matrices to obtain the spatial resolution mentioned at the beginning of this article. These electrodes can be surface electrodes, flexible electrodes, or electrodes penetrating the cortex and allowing its recording layer by layer (laminar electrode). The most significant electrodes that have been applied in vivo are (see **Figure 1**):

Neural-Matrix: an electrode matrix that can exceed one thousand recording electrodes (1008) with $195 \times 270$ μm² recording contacts spaced 250 to 330 μm apart on a 47.5 μm thick polyimide-elastomer matrix. The electrode density is 1212 per cm²[14].

Neuro-Grid: While this is not the only electrode using Pt-Au-PEDOT:PSS technology, it is one of the first published and most widely used[15]. It is a flexible electrode, 4 μm thick, composed of 256 recording channels of 10 μm$^2$ spaced by 30 μm and therefore having a spatial resolution of 111,000 electrodes per cm$^2$. The optimal distribution of the electrodes within the matrix for the Pt-Au-PEDOT:PSS in terms of density of size of exploration to obtain the most relevant information is an active area of research made possible by the ease of modifying the dimensions of the matrix and the distribution of the electrodes[16].

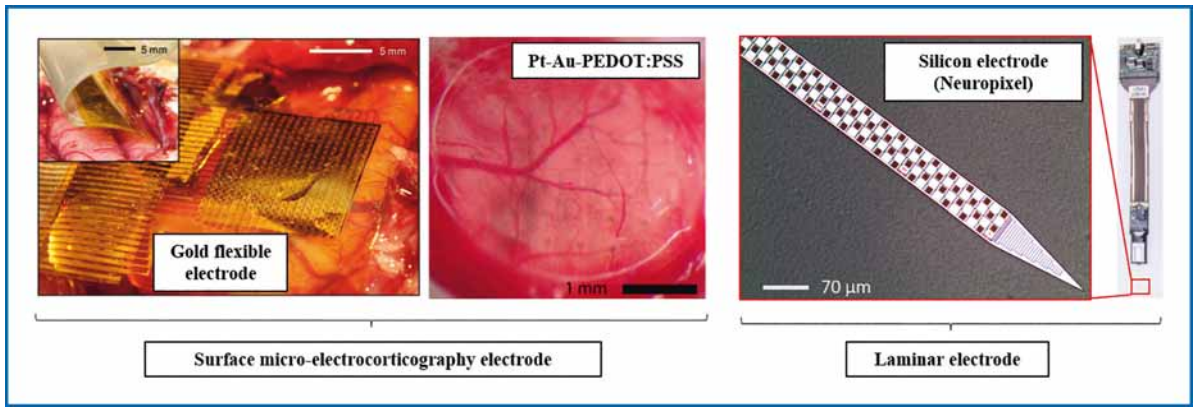

**Figure 1**: example of micro-electrocorticography (in gold[14] and Pt-Au-PEDOT:PSS[18]) and laminar (Neuropixel[19]) electrodes

Neuro-Pixel: unlike the two electrodes described above which are thin, flexible and applied to the surface of the cortex, this third type of electrode is called a laminar electrode, which penetrates the cortex to record from multiple layers. Where previous models had recording sites only at the end of the probe (such as the Utah Array recording only the third cortical layer), the recording electrodes on the Neuro-Pixel are distributed over the entire length of the probe, like the metallic laminar electrodes of the 20$^{th}$ century[17]. The Neuro-Pixel is 1cm × 70 μm × 20 μm, made of silicone, and has 12 μm$^2$ electrodes (between 250 to 1000 in number) made of titanium nitride with 15 μm spacing. The impedance is 0.2 Ωcm$^2$ for a frequency of sampling of 30 kHz.

The transparent electrodes used for optogenetics (mainly based on graphene or indium tin oxide) and whose prospect of human application is not immediate are not discussed here.

**Perspectives in medical research, description of a first approach**

The applications of these new technologies will undoubtedly be numerous in the field of medicine insofar as observing pathological processes on a scale that until now was not possible. The first application of these new electrodes that was deemed ethically acceptable in human primates was the case when it was already necessary to make a recording of the cortex, using older electrodes, for routine clinical procedures. In this case, a patient requires surgery as part of their treatment for drug-resistant epilepsy and the clinician seeks to identify the area of the brain responsible for generating epileptic seizures. The use of a new electrode in addition to the one used in the treatment adds only a minimal risk for the patient and has been considered acceptable by the appropriate ethics committee (Massachusetts General Hospital, Harvard Medical School). Thus, in 10 patients, one or more laminar-type electrodes were implanted and 30 epileptic seizures were recorded. Some electrodes were located outside the area generating epileptic seizures (but in areas that may be involved in their spread) and others within them. It was therefore possible to make comparisons between implantation sites. Analysis of this data highlighted a purely infra-granular and granular origin (i.e. the deepest layers of the neocortex) of the anomalies leading to epileptic seizure (see **Figure 2**). Conversely, in the areas of seizure propagation, it was noted that the generation of epileptic anomalies only concerned the granular layer, namely the most superficial layer of the neocortex (Bourdillon et al. Article under review). These results are unexpected because the normal organization of the neocortex follows a very robust vertical pattern, regardless of its location, in physiological units called the cortical column.

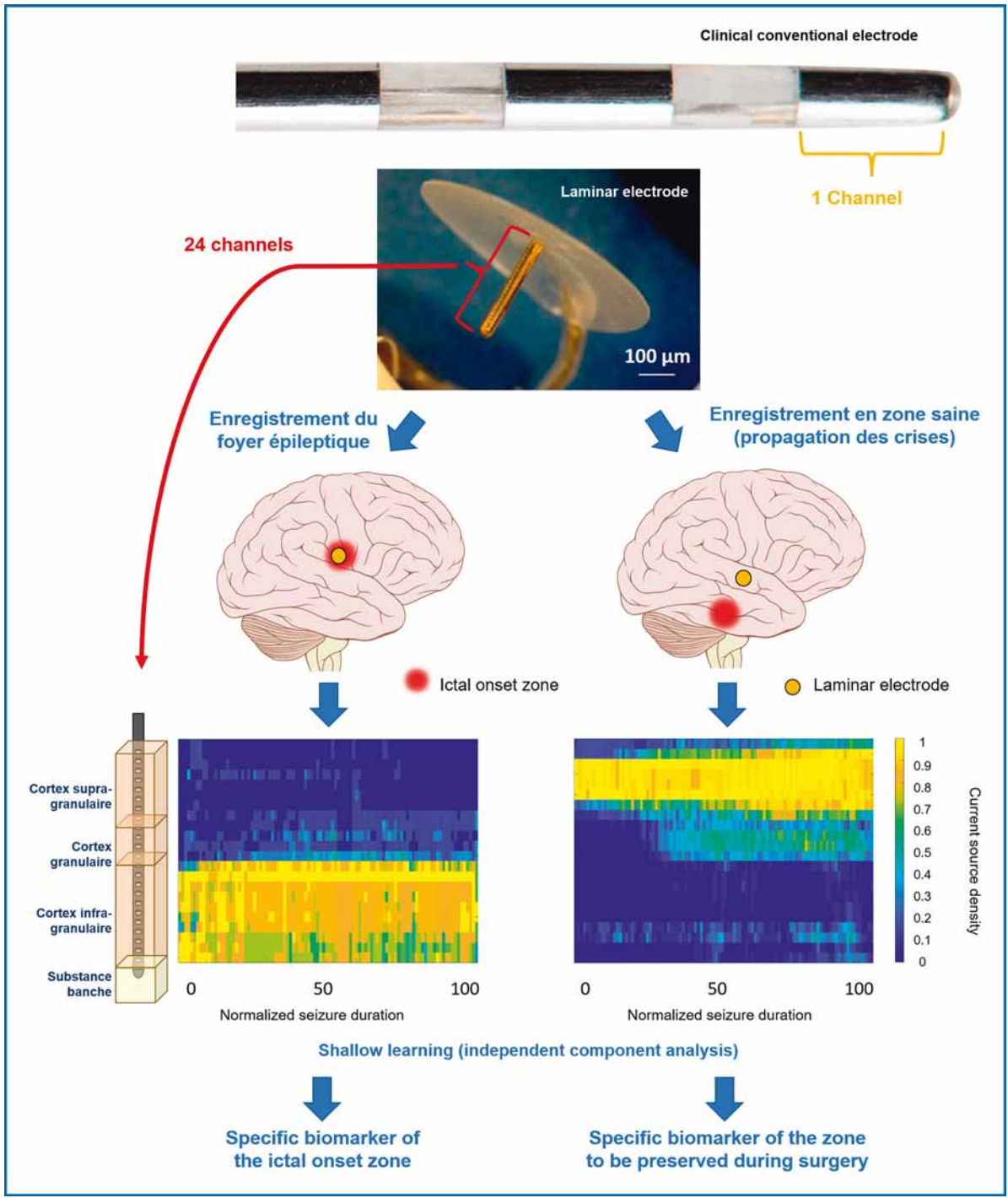

**Figure 2**: Graphical summary of the first results of laminar recordings carried out within the crisis generation focus as well as on its propagation path. The signal analysis is based on the calculation of the secondary spatial derivative of the field potentials (as known as current source density) allowing the estimation of the current flows between the different layers of the neocortex. The use of an independent component analysis allowed the identification of a specific biomarker of the focus of generation of seizures.

NEUROBIOLOGY

Indeed, the information arrives through the granular layer, and is then transmitted to the infragranular layer where an initial processing of the information is carried out before being projected, via pyramidal neurons, towards the most superficial layer of the granular cortex. A secondary processing of the information is then carried out, in particular comparing the information received with the prediction made locally, and the difference between these two elements is then projected remotely (to another cortical column) by pyramidal cells of the supra-layer granular (or sometimes deeper)[20,21].

The findings suggest a functional disconnect of the supra-granular layer from the cortical columns within the epileptic focus during the onset of an epileptic seizure. However, the mechanical electrodes used, through their 24 recording channels, did not make it possible to identify neurons on the basis of their multi-unit activity. A higher number and a denser distribution of electrodes would make it possible to obtain such information (for example to follow the activity of inhibitory interneurons, pyramidal cells, mirror cells of layer 6, and so on) and to identify the mechanisms responsible for this powerful vertical decoupling within a column. The NeuroPixel electrode offers this functionality and the first human recordings have now been made in a laboratory setting. However, only inter-critical signal is currently available and signal processing obstacles remain to be overcome (in particular the correction of motion artefacts)[22].

The use of laminar electrodes makes it possible to have a "vertical" view within a single cortical column but does not offer the possibility of analysing the interactions between the cortical columns in the same functional unit or at the same time within the epileptogenic zone (see **figure 3**). On the other hand, this is possible by using high resolution surface electrodes of the Pt-Au-PEDOT:PSS type, the spatial resolution, of the order of 30 to 50 μm depending on the configuration, being less than the size of a cortical column. The first human recordings have been made using these electrodes in the laboratory in epileptic patients, and enabled a high quality signal. However, no epileptic seizures have been recorded to date and all the data that has been analysed concerned recordings of interictal activity. Nevertheless it highlighted a pathological dynamic within the epileptic focus[16].

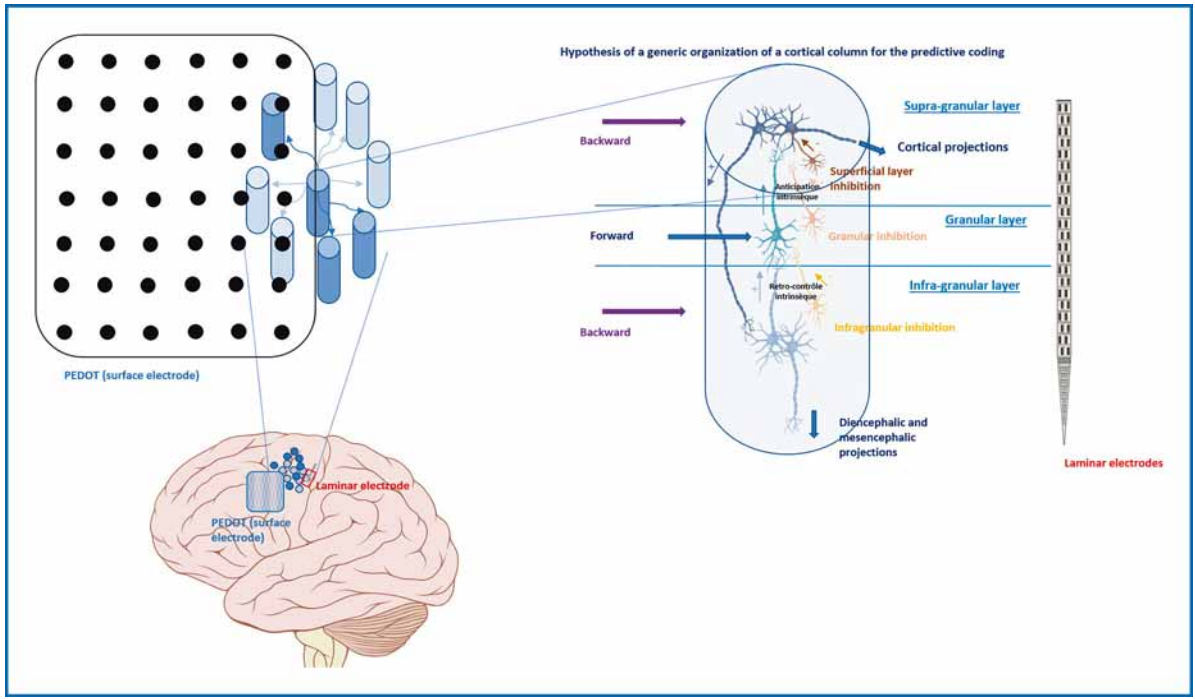

**Figure 3**: Example of recording on the one hand by Pt-Au-PEDOT:PSS allowing an analysis of a "horizontal" functional unit on the scale of the cortical columns, supposed seats of the elementary calculation operations of the neocortex, and on the other hand by a Neuropixel allowing "vertical" recording making it possible at the level of the neuron to have access to the cellular mechanisms involved in the realization of a calculation. The example taken here would test the so-called "predictive-coding" model proposed by Karl Friston[20], predicting that a cortical column is responsible for comparing the information received with a prediction made and then transmitting a quantification of the error observed. It would be a generic mechanism of information processing by the human neocortex. This hypothesis has not yet been tested.

### A perspective for understanding human cortical physiology

The possibility of making cortical recordings both from the surface over regions of several cm$^2$ and from the cortex in its depth with a resolution of less than 50 μm opens up the possibility of describing a still unknown cortical physiology in humans. Indeed, the in vivo use of these technologies in animals (mainly rodents, carnivores and non-human primates) has led to significant advances in understanding the basic mechanisms of brain functioning. Indeed, until now, the recordings were essentially based on a macroscopic recording of the brain and on the localization and temporality of events within large functional networks.

The possibility of observation on a more elementary scale (the neurons themselves within the smallest functional unit believed to exist within our brains) makes it possible to envisage the description of a still unknown cortical physiology in the human primate. Indeed, for these elementary processes of cortical functioning, there is a strong inter-species variability and the architecture of the human neocortex differs in many way from that of other species, including the non-human primate, in particular in the organization of layers 1 and 6, the latter being until now inaccessible to electrophysiological explorations. In addition, the human primate is characterized on the cognitive level by the possibility of constructing recursive reasoning, which seems essential for certain species-specific capacities such as the construction of language, mathematical structures or metacognition[23–25]. In addition to understanding and studying the assumptions made about the basic functioning of the brain as a calculator making inferences and quantifying prediction errors[20,22,23] (see **Figure 3**), it is becoming possible to make inter-species comparisons and to isolate the process or set of processes allowing the human species to produce recursive reasoning.

### Conclusion

The development of new materials for the design of electrodes which allow recording of the brain at the scale of the elementary functional units (the cortical columns and the neurons composing them) makes it possible to address questions which until now could not be answered in terms of cortical physiology and pathophysiology. In a first work we have proven this on the very precise question of the mechanism of the generation of epileptic seizures and have succeeded in isolating a specific biomarker opening the way to a concrete neurosurgical application in the very short term (electrophysiological guidance of resection of the epileptic focus). We have also demonstrated the feasibility of recordings with all the techniques proposed. In addition to the possible advances in the field described previously, it is also possible that these electrodes allow decoding of neural activity with high quality, and chronically implanted electrodes could be used for brain machine interfaces, particularly with a view to compensate for impaired functions following damage to the central nervous system.

> "The possibility of making cortical recordings both from the surface over regions of several cm$^2$ and from the cortex in its depth with a resolution of less than 50 μm opens up the possibility of describing a still unknown cortical physiology in humans."

### Acknowledgements

Pr Sydney S Cash, Drs Angelique Paulk & Pariya Salami from Harvard Medical School – Massachusetts General Hospital; Mila Haglren from Massachusetts Institute of Technology; Pr Eric Hagren from University of California San Diego; Société Française de Neurochirurgie; Fulbright Monahan grant; Edmond de Rothschild fundations.